\documentclass[twocolumn]{aastex701}

 \usepackage{amsmath, float, bbm} 
 \usepackage{tabularx}

\begin{document}


\title{An Open Cluster Origin for Most Solar-Type Binaries in the Solar Neighborhood and Implications for Primordial Planet Stability}

\author[0000-0002-9343-8612]{Anna C. Childs}
\affiliation{Center for Interdisciplinary Exploration and Research in Astrophysics (CIERA) and Department of Physics and Astronomy Northwestern University,
1800 Sherman Ave, Evanston, IL 60201 USA}

\affiliation{Department of Physics and Astronomy, The University of Alabama, Box~870324, Tuscaloosa, AL~35487-0324, USA}
\email{acchilds1@ua.edu}

\author[0000-0002-3881-9332]{Aaron M. Geller}
\affiliation{Center for Interdisciplinary Exploration and Research in Astrophysics (CIERA) and Department of Physics and Astronomy Northwestern University,
1800 Sherman Ave, Evanston, IL 60201 USA}
\email{a-geller@northwestern.edu}



\begin{abstract}
We present the first Gaia-calibrated estimate of the contribution of dynamically processed open clusters (OCs) to the nearby solar-type field binary population and investigate the implications for primordial planet stability.  We combine \textit{Gaia} DR3 observations, empirical models of cluster formation and dissolution, and direct $N$-body simulations to constrain the fraction of field FGK-type binaries within $1\,\mathrm{kpc}$ that originated in OCs over the past $5\,\mathrm{Gyr}$. We find that at least $\sim 53\%$ of solar-type field binaries formed in OCs with initial masses $\geq200\,M_{\odot}$, where dynamical processing becomes significant. We quantify the effects of stellar encounters and binary orbital evolution on the stability of both circumbinary and circumstellar planets by deriving empirical prescriptions for circumbinary planet destabilization as a function of binary separation and for the fractional reduction of stable circumstellar coplanar phase space. These results provide a framework for evaluating how OC dynamical evolution may influence the stability of planets in binary systems and therefore contribute to observed exoplanet demographics.

\end{abstract}


\section{Introduction} \label{sec:intro}

Most stars are thought to form in stellar clusters and associations embedded within giant molecular clouds before dispersing into the Galactic field \citep{Pudritz2002, Lada2003, Pfalzner2011, Quintana2025}. While many birth environments dissolve shortly after the embedded phase, denser systems can survive as open clusters (OCs), spanning a wide range of masses, ages, and metallicities.

Binary stars are common both in OCs and in the field, where solar-type stars exhibit a binary fraction of roughly $50\%$ that increases with primary mass \citep{Raghavan2010, sana12, Duchene2013, cab14}. Observed OC binary fractions vary substantially with cluster age and mass \citep{Jadhav2021, Cordoni2023, Donada2023, Pang2023, Childs2025, Malhotra2026}. Stellar encounters within OCs can significantly modify binary orbits before the systems are ejected into the field. According to Heggie’s law, wide (``soft") binaries generally widen further while tight (``hard”) binaries become more tightly bound through repeated encounters \citep{Heggie1975, Hills1975}. These orbital changes alter the regions where planets can remain stable. Consequently, present-day binary architectures may not accurately reflect the environments in which their planets formed.

Numerous studies have examined planet stability and survival around single stars in OCs \citep{Laughlin1998, Adams2001, Scally2001, Adams2006, Malmberg2007, Proszkow2009, Li015, Zheng2015, Cai2017, Fugii2019, Hands2019, Li2019, vanElteren2019, Daffern-Powell2022, Parker2026, Webb2026}.  These calculations have demonstrated a wide range of possible dynamical outcomes, including orbital excitation, planet–planet scattering, ejection, exchange between stellar hosts, and the capture of free-floating planets.

Comparatively few have considered planets in binary systems despite observational evidence that OC binaries are preferentially mass segregated and therefore experience enhanced encounter rates \citep{Milone2012, Jadhav2021, Motherway2024, Zwicker2024, Childs2025}. Using direct $N$-body simulations, \citet{Liu2013} found that circumstellar planets are less likely to survive in binaries, although systems with periapsis distances exceeding $100 \,\rm au$ generally retain their planets. More recently, \citet{Li2024} showed that stellar encounters in OCs efficiently trigger von Zeipel–Lidov–Kozai evolution capable of producing hot Jupiters \citep{vonzeipel1910, Kozai1962, Lidov1962, Ito2019}.

Planet occurrence rates in binary systems remain less well constrained than those around single stars because of observational biases, although measurements now exist for both circumstellar and circumbinary planets (CBPs). For binaries with orbital periods less than $300 \, \rm days$, the occurrence rate for coplanar CBPs with radii of at least $6 \, R_\oplus$ is $10.0^{+18}_{-6.5}\%$ \citep{Armstrong2014}. Circumstellar planet occurrence rate has been found to increase with binary separation, but overall an occurrence rate of $0.12 \pm0.4$ planets per star was found for planets with masses in the range of $0.1-10 \, M_{\rm J}$ \citep{Hirsch2021}. Recent work also suggests that rocky planets are intrinsically less common in binary systems than around single stars \citep{Sullivan2026}. However, these occurrence rates have generally not been interpreted in the context of the dynamical evolution experienced by binaries in their natal clusters.

We combine $N$-body simulations spanning a range of cluster environments with observationally motivated models of OC formation and dissolution to estimate the contribution of dynamically processed binaries to the field population within $1 \,\rm kpc$ over the past $5 \,\rm Gyr$. We calibrate our results using recent OC catalogs from Gaia. Lastly, we use the simulated binary orbital histories together with stellar encounter timescales to quantify the probabilities of the primordial circumstellar and CBP stability.  

Throughout this work, we use the term \textit{primordial planet} to refer to a planet that formed with and remains associated with its parent stellar or binary system, distinguishing these planets from those subsequently acquired through dynamical capture or exchange.  We refer to OCs with initial masses $\ge200\,M_\odot$ as \textit{dynamically processed OCs}. 

Section~\ref{sec:methods} describes the simulations and methodology. Section~\ref{sec:occurrence_rates} presents our results, and Section~\ref{sec:conclusions} summarizes our conclusions.

\begin{figure*}
\includegraphics[width=.5\textwidth]{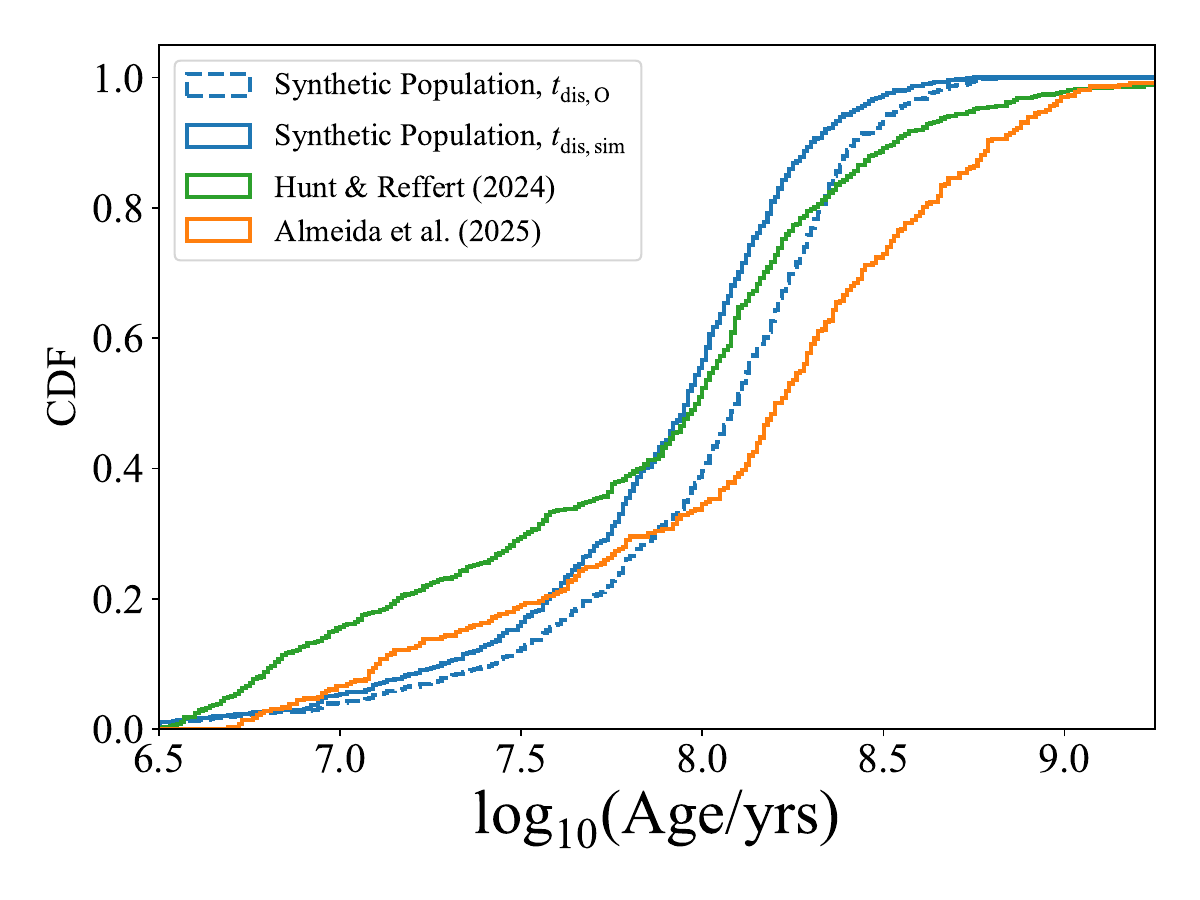}
\includegraphics[width=.5\textwidth]{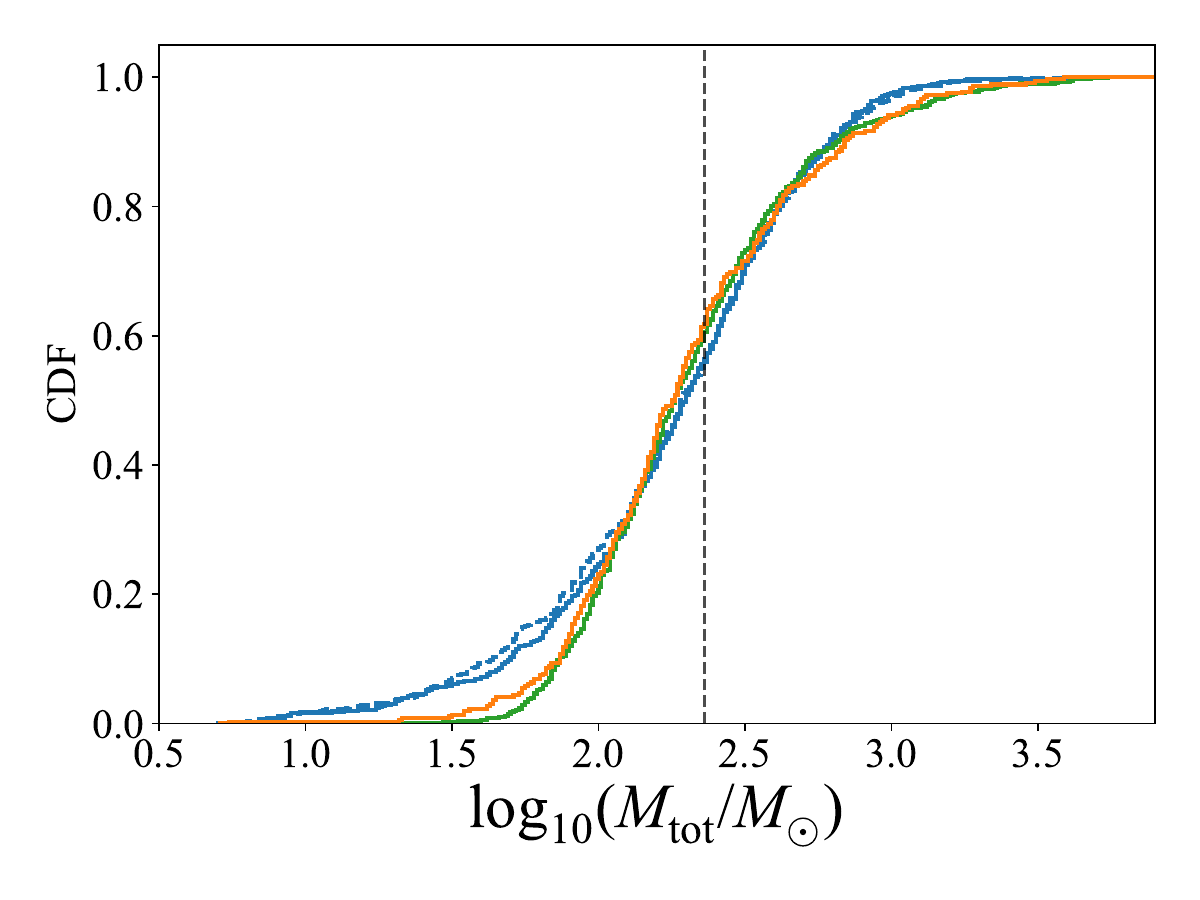}
    \caption{Age (left) and mass (right) CDFs for the surviving OCs, no mass cuts imposed, in one realization of our synthetic population.  We compare these distributions to the observations of all OCs within $1 \, \rm kpc$ from the \cite{Hunt2024} and \cite{Almeida2025} catalogs with at least a mass of $230 \, M_{\odot}$ (marked by the dashed vertical line on the right panel) where observations are expected to be complete.  The blue dashed line shows our synthetic extant OC distributions using $t_{\rm dis,O}$ and the solid blue line shows these distributions using $t_{\rm dis,sim}$.}
    \label{fig:age_mass_hist}
\end{figure*}

\section{Methods}\label{sec:methods}
We use the publicly available stellar $N$-body code \texttt{nbody6++GPU} to simulate stellar evolution and the gravitational interactions of stars in an OC \citep{Wang2015code}.  Although this implementation supports GPU acceleration, we made use of the CPU parallelization of this code for all of the reported simulations.  We simulate the evolution of 235 OCs up until cluster dissolution.  All OCs are initialized with a solar galactic distance, a solar-type metallicity ($Z=0.02$), and a range of initial half-mass radii.  We vary the initial number of stellar systems (single and binary stars) that an OC begins with, spanning from 500 to 40,000 and simulate the evolution of multiple realizations for each setup (each with a different initial random seed). The initial stellar masses are randomly drawn from a standard \citet{kroupa2001} initial mass function (IMF).  For a given binary system, the two component masses are first combined, and then split according to the selected mass-ratio to produce the final component masses for the system. 

Stellar positions and velocities are randomly sampled from the equilibrium Plummer distribution function. Individual realizations exhibit finite fluctuations about the idealized equilibrium configuration, but all models are initialized in approximate virial equilibrium according to a Plummer sphere \citep{Plummer1911}.  A smooth, virialized Plummer model does not capture the substructure and non-equilibrium conditions of some young clusters, which can enhance early dynamical processing \citep{Goodwin2004, Rossi2017}. Our models may therefore underestimate this early processing, although the influence of these initial conditions diminishes as clusters relax toward smoother configurations \citep{Geller2013b}.

In all OCs the initial binary fraction is a function of binary primary mass, in an attempt to model the trend observed on the field by \citet{Raghavan2010} and also in OCs \citep[e.g. see][and references therein]{Childs2024, Childs2025, Fuhrmass2017, Kaczmarek2011, Malhotra2026,  sana12}, where solar-type stars have a binary fraction of about 50\%, and this increases (decreases) as the primary mass increases (decreases).  The adopted initial binary fractions are also consistent with the binary fraction measured for thin-disk G- and K-dwarf field stars by \citet{Niu_2021}. The binary mass-ratio, eccentricity and period values are random draws from observed distributions; for primary masses less than $15 \, M_{\odot}$, we draw from distributions observed by \citet{Raghavan2010} while for O stars, we substitute the period distribution from \citet{sana12}.  

All stars begin as main-sequence (MS) stars and proceed through stellar evolution processes as the simulation time progresses.  Mass transfers and dynamical kicks from supernovae are included in these models. The simulations use the single star and binary stellar evolution prescriptions which are based on the algorithms of \cite{Hurley2000, Hurley2002}, respectively.

New binaries can be formed through dynamical interactions. However, we find in all our simulations that the destruction of soft binaries is dominant over dynamical formation pathways and the final number of binaries in an OC is always lower than what it was initialized with.  We do not include any planets in our simulations. Consequently, our calculations do not explicitly follow planet-level dynamical outcomes such as scattering, ejection, collision, exchange between stellar hosts, or the capture of free-floating planets. Instead, we follow the dynamical evolution of the stellar binaries themselves and subsequently evaluate how encounter-driven changes to their orbital properties modify the stable phase space available to primordial planets. 

All OCs are integrated until they have ten objects (stars or binaries) left in the OC.  However, multiple escapers can be removed during a single escape-processing step and the final number of objects can fall below ten before termination.  A star or binary is defined as escaped when it reaches a distance greater than twice the tidal radius from the density center of the OC.


\begin{table}[t]
\centering

\begin{tabular}{|c|c|c|}
\hline
Parameter & $t_{\rm dis, O}$ & $t_{\rm dis, sim}$ \\
\hline
CFR ($\rm Myr^{-1} \,  kpc^{-2}$) & 0.86 & 1.25 \\
\hline
Total no. of OCs & 13,508 & 19,634  \\
\hline
No. of extant OCs & $297 \pm 10$ & $299 \pm 13$ \\
\hline
\% all binary candidates & $52.8\pm0.3\%$ &  $75.6\pm0.5\%$ \\
\hline
\% high-prob. candidates & $85.1\pm0.5\%$ & $100\pm0.9\%$ \footnote{The inferred fraction exceeds 100\%; we report 100\% because the high-probability sample is likely incomplete.} \\
\hline

\hline
\end{tabular}
\caption{The resulting cluster formation rate (CFR), total number of OCs formed over the last $5 \, \rm Gyr$, number of extant OCs, and the fraction of all binaries and high-probability binary candidates from \cite{ElBady2021} that have originated from a dynamically processed OC.  Each row lists these values for the two different dissolution times we investigate. The uncertainties on these values come from our bootstrapping process.}
\label{tab:dis_times}
\end{table}

\subsection{Weighting the Simulations}
To constrain the occurrence rates of ejected binaries, we weight our $N$-body simulations against a synthetic population of surviving and dissolved OCs generated from observationally motivated cluster formation and dissolution models. We model the OC dissolution time, $t_{\rm dis}$, using both the measured dissolution times from our simulations and literature estimates. Combined with an initial cluster mass function (ICMF), these models determine the cluster formation rate (CFR) required to reproduce the observed number of OCs within $1\,\rm kpc$ with masses above $230\,M_\odot$, where observations are expected to be complete. We then generate a synthetic OC population over the past $5\,\rm Gyr$, map each cluster onto our simulation grid, and estimate the number of FGK MS+MS binaries in the solar neighborhood that we expect to have originated from OCs with masses of at least $200 \, \rm M_{\odot}$.

The dissolution time of an OC with initial mass $M_{\rm i}$ is commonly approximated as
\begin{equation}\label{eq:dis_time}
    t_{\rm dis}=\gamma t_{4}^{\rm tot}(M_{\rm i}/10^4M_{\odot})^{\gamma},
\end{equation}
where $t_{4}^{\rm tot}$ is the dissolution timescale associated with the combined effects of the galactic tidal field, stellar evolution, spiral-arm shocking, and molecular cloud encounters, and $\gamma$ is a free parameter that depends on the concentration of stellar mass within the OC \citep{Almeida2025, Boutloukos2003,Lamers2005}.

The corresponding mass-loss rate is
\begin{equation}\label{eq:mass_loss_rate}
    \frac{\mathrm{d}M}{\mathrm{d}t}= - \frac{(M_{\rm i})^{1-\gamma}(10^4)^{\gamma}}{t_{4}^{\rm tot}\gamma},
\end{equation}
\citep{Almeida2025}.

Using Gaia DR2 OCs within $1.5\,\rm kpc$, \citet{Almeida2025} found that a skewed log-normal ICMF with $\alpha=2$, location $=2.3$, scale $=0.5$, together with $t_4^{\rm tot}=3\,\rm Gyr$ and $\gamma=0.6$--$0.7$, reproduces the observed cluster population. We find that this ICMF overpredicts the masses of surviving OCs and instead adopt a scale parameter of $0.35$, which better matches the age and mass distributions reported by \citet{Almeida2025} and \citet{Hunt2024}. Throughout this work, the corresponding observational dissolution model uses $\gamma=0.6$ and $t_4^{\rm tot}=3\,\rm Gyr$ and is denoted $t_{\rm dis,O}$.

For each synthetic OC, we draw an age uniformly between $0$ and $5\,\rm Gyr$ and an initial mass from the adopted ICMF, then compute its dissolution time. Clusters with $t_{\rm dis}$ shorter than their assigned age are classified as dissolved; otherwise they survive to the present day. For surviving OCs, we calculate the present-day mass using Equation~\ref{eq:mass_loss_rate}. We generate ten realizations with different random seeds to estimate statistical uncertainties.

\citet{Hunt2024} identified 6956 bound OCs in Gaia DR3 with completeness-corrected photometric masses and showed their catalog is complete for $M_{\rm now}\ge230\,M_\odot$ within $1.8\,\rm kpc$. Restricting this catalog to $1\,\rm kpc$ yields 298 extant OCs, which we adopt as the observational benchmark for calibrating our synthetic populations.

Assuming a constant cluster formation rate (CFR) over the last $5 \,\rm Gyr$, we find that a CFR of $0.86 \,\rm Myr^{-1} \,kpc^{-2}$ reproduces the observed number of extant OCs when using the observationally motivated dissolution model, $t_{\rm dis,O}$. This value is within $2\sigma$ of the present-day CFR derived by \citet{Anders2021}, who found $0.55^{+0.19}_{-0.15} \,\rm Myr^{-1} \, kpc^{-2}$ from the age distribution of Gaia DR2 OCs within a $2 \,\rm kpc$ cylinder centered on the Sun. We note that a constant CFR is unlikely, but more work is needed to better constrain a variable CFR in the past.

The dissolution timescale of an OC also depends on both its IMF and initial binary fraction \citep{Haghi_2020}. Because these parameters are not explicitly included in Eq.~\ref{eq:dis_time} we also determine the dissolution timescale directly from our $N$-body simulations, which self-consistently include stellar evolution and span a range of initial binary fractions. We note, however, that all of our simulations adopt the standard \citet{kroupa2001} IMF and are placed at the solar galactic radius, thus experiencing the same galactic potential which affects tidal stripping efficiency. Fitting Eq.~\ref{eq:dis_time} to the simulated dissolution times yields $\gamma=0.67$ and $t_4^{\rm tot}=2.25\,\rm Gyr$, defining $t_{\rm dis,sim}$. This model requires a larger best-fit CFR of $1.25\,\rm Myr^{-1}\,\rm kpc^{-2}$ to reproduce the observed OC population.

The uncertainty in the OC dissolution timescale translates directly into uncertainty in the number of clusters that have formed over the last $5 \,\rm Gyr$. Consequently, the inferred contribution of OCs to the field population is sensitive to the adopted dissolution model. To quantify this uncertainty, we perform all subsequent analyses using both $t_{\rm dis,O}$ and $t_{\rm dis,sim}$. A summary of the resulting parameters is provided in Table~\ref{tab:dis_times}.

Figure~\ref{fig:age_mass_hist} compares the age and mass distributions of surviving synthetic OCs (no mass cuts) from one realization with the observed populations from \citet{Almeida2025} and \citet{Hunt2024}. Solid and dashed blue curves correspond to the $t_{\rm dis,sim}$ and $t_{\rm dis,O}$ models, respectively. Both models reproduce the observed distributions well.  

\begin{figure}
\includegraphics[width=.5\textwidth]{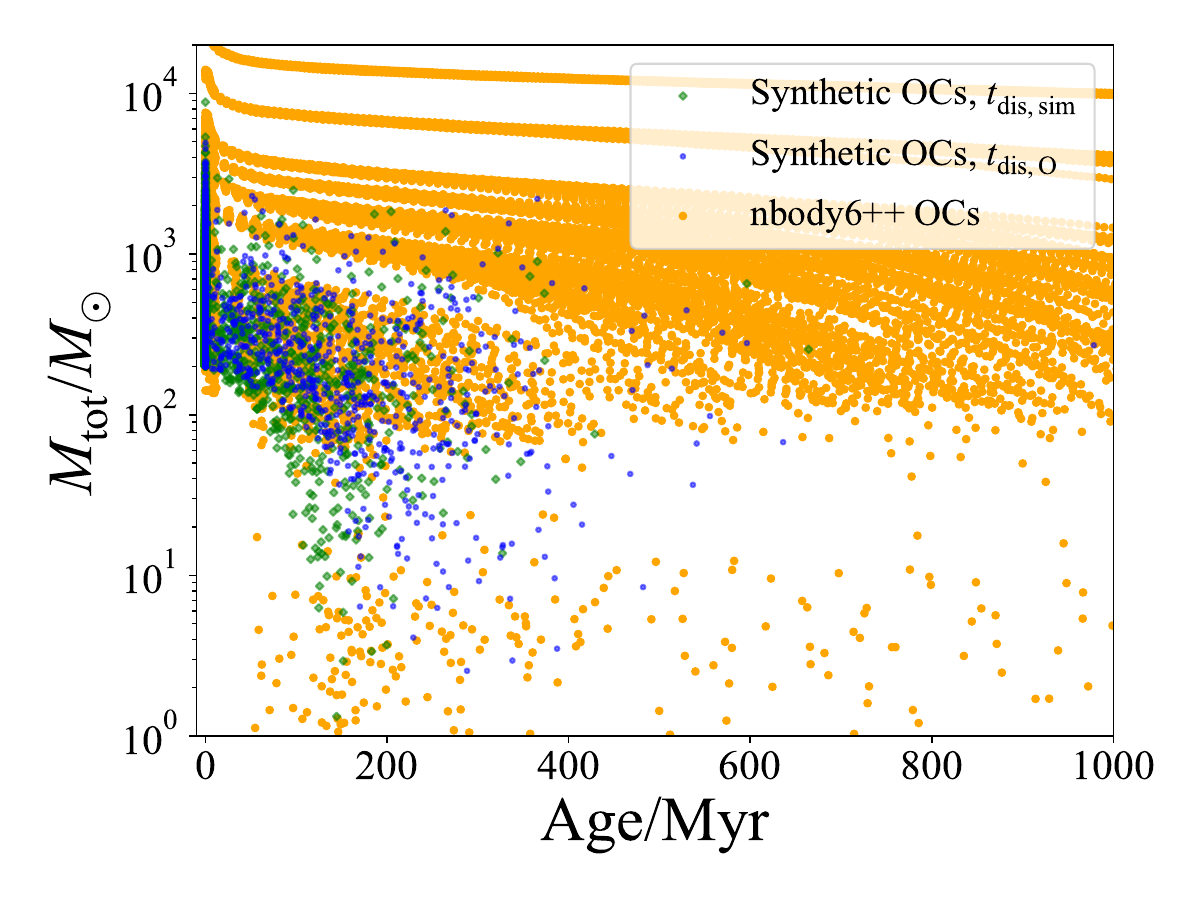}
\caption{The synthetic OCs generated over $5 \rm \, Gyr$ within $1 \, \rm kpc$ with $t_{\rm dis, O}$ and $t_{\rm dis, sim}$ are shown in blue dots and green diamonds, respectively.  We show the initial mass at age = 0 for all of the dissolved OCs over the last $5 \, \rm Gyr$ and we show the current mass and age of the surviving OCs.  All recorded snapshots from our simulations are shown with orange dots.
}
    \label{fig:draws}
\end{figure}

We match the surviving synthetic OCs to the nearest $N$-body simulation snapshot in present-day age and mass using a \texttt{scipy cKDTree} \citep{ckdtree199, SciPy}. We match dissolved OCs to simulations by the OC total mass at $1\,\rm Myr$, avoiding biases from OCs initialized slightly out of equilibrium that may have experienced rapid early mass loss.  Figure~\ref{fig:draws} compares one realization of the synthetic population with the \texttt{nbody6++} simulation grid. The $t_{\rm dis,O}$ and $t_{\rm dis,sim}$ models produce 13,508 and 19,634 OCs over $5\,\rm Gyr$ within $1\,\rm kpc$, respectively. Dissolved clusters are plotted at their initial masses ($t=0$), while surviving clusters are shown at their present-day ages and masses. The synthetic populations are well covered by the simulation grid.

\begin{figure*}
\includegraphics[width=.5\textwidth]{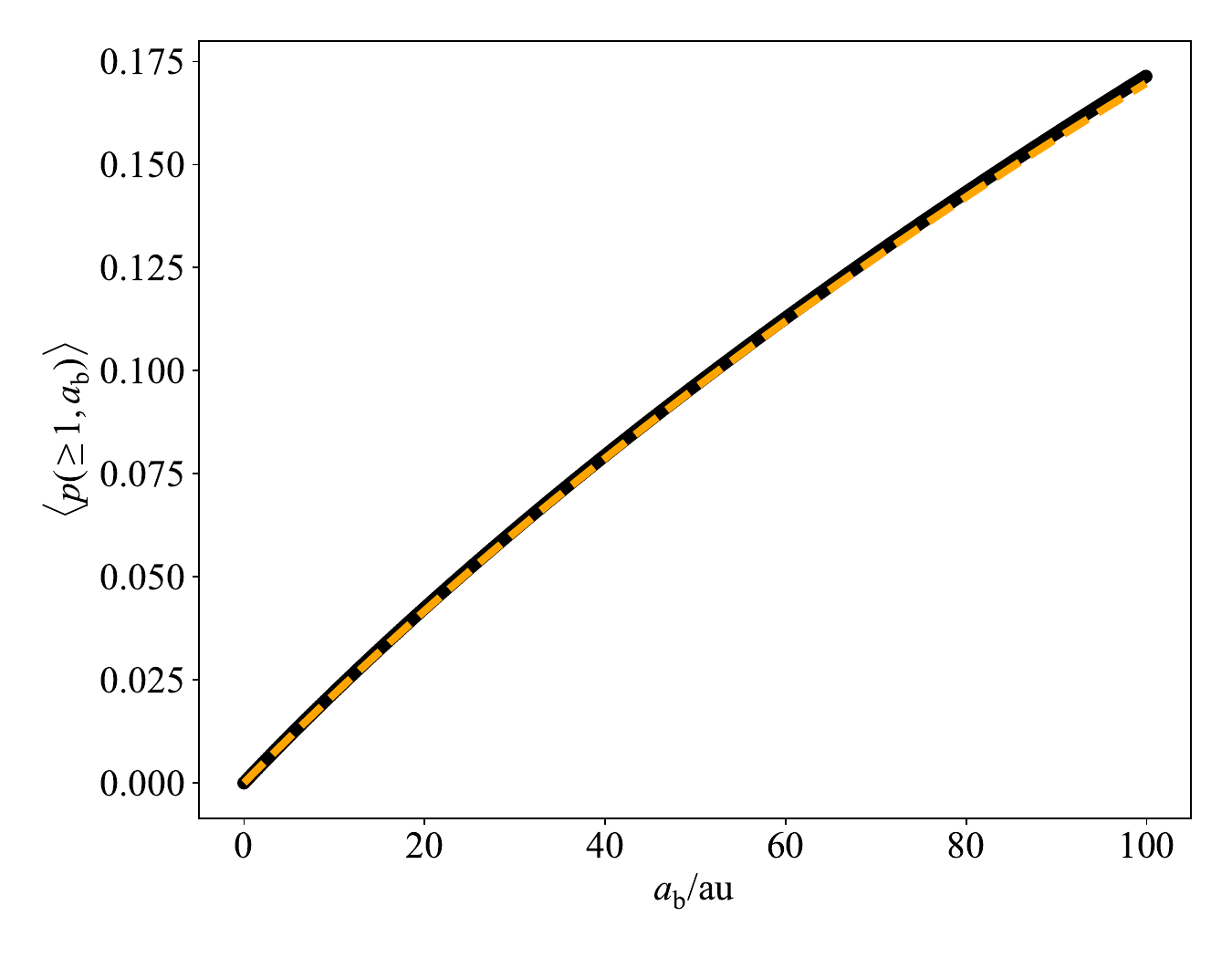}
\includegraphics[width=.5\textwidth]{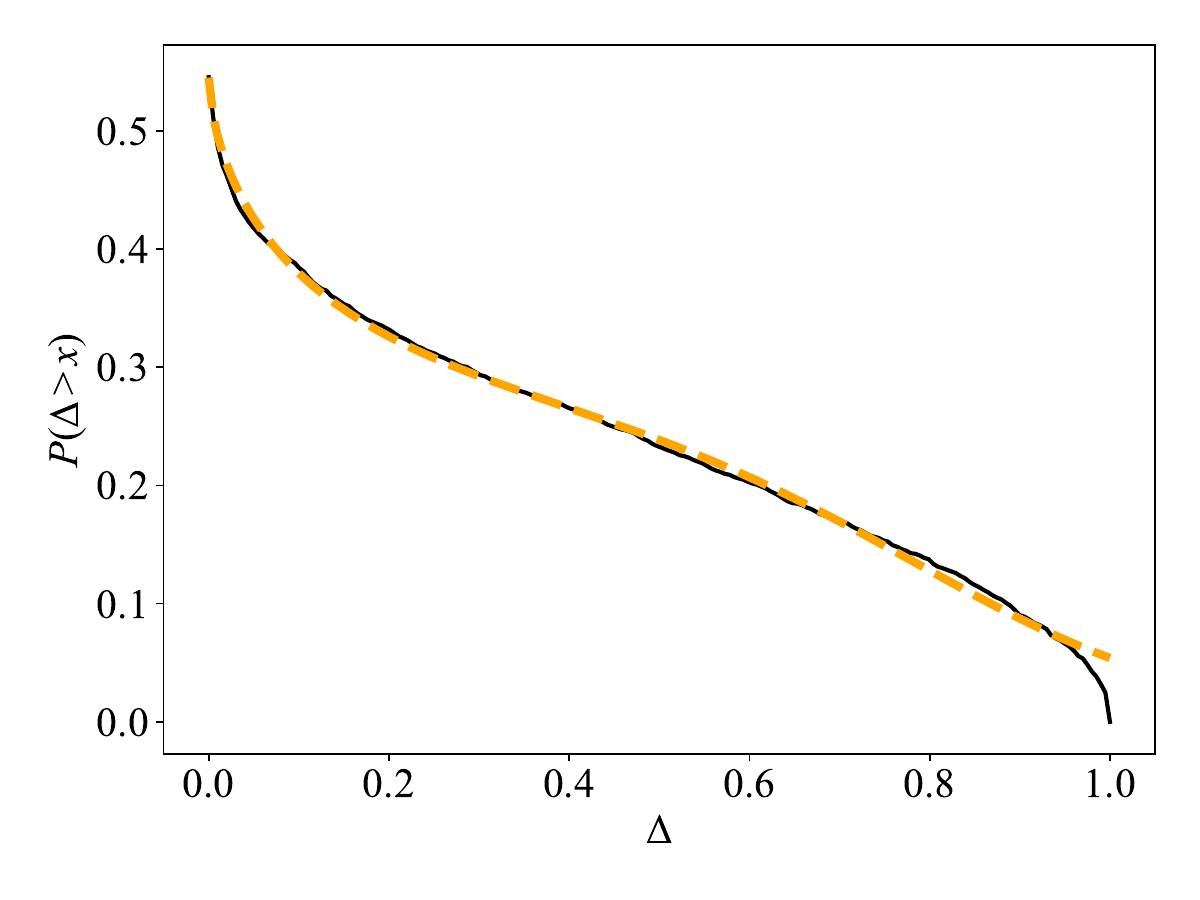}

    \caption{Left: Calculated $\langle p(\geq 1, b_{\rm gf}) \rangle$, the probability of a binary with a given $a_{\rm b}/\rm au$ experiencing a CBP destabilizing stellar encounter in its natal OC.  Right: Empirical stability curve of primordial circumstellar planets $P(\Delta > x)$. The data are shown in black and the fits are shown by the orange dashed lines.}
    \label{fig:prob_fit}
\end{figure*}

\section{Occurrence Rates and Primordial Planet Stability}\label{sec:occurrence_rates}
Using our $N$-body results, our synthetic OC population, and observations of binaries in the field, we calculate the occurrence rate of FGK binaries in the field that have originated from an OC over the last $5 \, \rm Gyr$. Additionally, we quantify how OC dynamical processing reduces the stable phase space available to primordial circumbinary and circumstellar planets in these systems.

\subsection{Field binaries from OCs}
Using \textit{Gaia} EDR3, \citet{ElBady2021} identified 1,059,677 candidate MS+MS binaries with separations below $1\,\rm pc$ within $1\,\rm kpc$ and primary masses of $0.4<M_1/M_{\odot}<1.3$, including a high-probability subsample of 658,062 binaries. This sample has been cleaned to remove clusters, moving groups, background pairs, and triples.

To estimate the fraction of these binaries that originated from OCs with initial masses $\ge200\,M_{\odot}$ over the past $5\,\rm Gyr$, we count MS+MS binaries in the same primary-mass range that escaped from our \texttt{nbody6++} simulations and were matched to synthetic OCs. For surviving clusters, we include escapers from $1\,\rm Myr$ to the cluster's present age. For dissolved clusters, we include all escapers between the matched starting point and cluster dissolution.  We consider both primordial and new binaries (binaries formed through dynamical encounters and exchanges) in these counts.

Table~\ref{tab:dis_times} summarizes the results for both dissolution models and both \citet{ElBady2021} binary samples. We find that $\sim 53$--$100\%$ of solar-type field binaries within $1\,\rm kpc$ originated from dynamically processed OCs over the past $5\,\rm Gyr$, with more than $\sim90\%$ contributed by clusters that have since dissolved. The $100\%$ value we report is most likely the result of incompleteness in the \citet{ElBady2021} high-probability subsample, especially for wide binaries.

These results are consistent with predictions that most field stars formed in embedded clusters, the vast majority of which subsequently dissolved \citep{Lada2003}, as well as the more recent estimate that roughly $50$--$80\%$ of solar-neighborhood field stars originated in OCs \citep{Quintana2025}.  Additionally, an analysis of solar twins within $300 \, \rm pc$ from Gaia DR3 data found the age distribution to have a narrow peak at $2 \, \rm Gyr$ and a broader peak over $4-6 \, \rm Gyr$ \citep{Tsujimoto2026} and \citet{Mor2019} found evidence in Gaia DR2 data of a burst in star formation rate in the thin-disk $2-3 \, \rm Gyr$ ago.  While these epochs of star-burst formation are not consistent with a constant CFR we use in our models, we note the estimated age range of nearby solar type binaries, along with the number of observed extant OCs, is consistent with our results.

\subsection{Circumbinary planets from OCs}\label{sec:CBPs}

To quantify the impact of OC dynamics on CBPs we calculate the expected number of stellar encounters capable of destabilizing a CBP. We approximate the binary--binary encounter timescale using Eq. A8 of \citet{Leigh2011},
\begin{align}\label{eq:encounter_time}
\tau_{2+2} = 1.3 \times 10^7\, f_b^{-2} \left( \frac{1\, \mathrm{pc}}{r_{\rm c}} \right)^3 \left( \frac{10^3\, \mathrm{pc}^{-3}}{n_{0}} \right)^2 \notag\\
\times
\left( \frac{v_{\mathrm{rms}}}{5\, \mathrm{km\,s}^{-1}} \right)
\left( \frac{0.5\, M_\odot}{\langle m \rangle} \right)
\left( \frac{1\, \mathrm{au}}{a_{\rm b}} \right)\ \mathrm{yr},
\end{align}
where $f_{\rm b}$ is the binary fraction, $r_{\rm c}$ is the core radius, $n_0$ is the central number density, $v_{\rm rms}$ is the velocity dispersion, $\langle m \rangle$ is the mean stellar mass, and $a_{\rm b}$ is the binary separation. We estimate
\begin{equation}
n_0=\frac{3N_{\rm core}}{2\pi r_c^3},
\end{equation}
following \citet{King1966}, where $N_{\rm core}$ is the number of objects within the 3D density-weighted core radius returned by \texttt{nbody6++}. We adopt the simulated $N$-body cluster properties at each snapshot, assume $v_{\rm rms}=1\,{\rm km\,s^{-1}}$, and leave $a_{\rm b}$ as a free parameter.

We adopt binary--binary encounters because they dominate over single--binary encounters in OC cores when $f_b\gtrsim0.25$ \citep{Leigh2011}, with mass segregation expected to further enhance their importance \citep{Childs2025}. Since the cluster properties evolve with time, we evaluate Eq.~\ref{eq:encounter_time} at every simulation snapshot. We estimate the cumulative number of encounters as
\begin{equation}\label{eq:Nenc}
N_{\rm enc}=\sum_j \frac{\Delta t_j}{\tau_j},
\end{equation}
where $\tau_j$ is the encounter timescale at snapshot $j$ and $\Delta t_j$ is the elapsed time since the previous snapshot.

Assuming encounters follow Poisson statistics, the probability of one or more encounters is
\begin{equation}
P(\ge1)=1-e^{-N_{\rm enc}}.
\end{equation}
To estimate the probability that a randomly chosen binary undergoes such an encounter across all simulations, we compute a weighted average where we weight each $P(\ge1)$ for a given simulation by the number of binaries it contributes to the field population.

Because $\tau_{2+2}\propto a_{\rm b}^{-1}$, the expected number of encounters scales linearly with binary separation. We therefore calculate the weighted probability for $a_{\rm b}\in[0,100]\,{\rm au}$ and fit the results with
\begin{equation}\label{eq:fit}
f(a_{\rm b})=0.3273\,\ln\!\left(1+0.0068\,a_{\rm b}\right).
\end{equation}

The left panel of Figure~\ref{fig:prob_fit} compares the calculated probabilities with this fit. Equation~\ref{eq:fit} can be used to estimate the probability that a binary experienced a CBP-destabilizing encounter while it was embedded in its natal OC. For example, $\alpha$ Centauri AB ($a_{\rm b}\approx23\,{\rm au}$) has a predicted probability of $\sim5\%$, whereas 61 Cygni AB ($a_{\rm b}\approx87\,{\rm au}$) has a probability of $\sim15\%$.  Neither of these systems currently have confirmed CBPs, but these values quantify the likelihood that each system experienced at least one stellar encounter in an OC that would destabilize any formed CBP.

These estimates assume coplanar CBPs. Polar circumbinary planets can remain stable at smaller orbital distances than coplanar planets \citep{Childs2021}, suggesting they may be more resilient to stellar flybys. Given that a substantial fraction of circumbinary disks are predicted to evolve into polar configurations \citep{Johnson2025}, future work should investigate how planetary inclination modifies encounter stability probabilities.

\subsection{Coplanar circumstellar phase space}
Section~\ref{sec:CBPs} showed that the probability of a stellar encounter in an OC increases with binary separation yet circumstellar planet occurrence rates increase with binary separation \citep{Hirsch2021}. Consequently, circumstellar planet occurrence reflects a trade-off between formation efficiency and dynamical stability.

To quantify this effect, we follow the orbital evolution of each simulated primordial FGK binary with $a_{\rm b}\ge100\,{\rm au}$ at $t=1\,{\rm Myr}$. Using the S-type stability criterion of \citet{Holman1999} for coplanar circular orbits, we compute the initial stable circumstellar radius, $a_{\rm c,0}$, and the minimum stable radius reached before escape from the cluster, $a_{\rm c,min}$. We define the fractional reduction in stable circumstellar phase space as
\begin{equation}
\Delta = 1-\frac{a_{\rm c,min}^2}{a_{\rm c,0}^2}.
\end{equation}

The median is $\Delta=0.029^{+0.760}_{-0.029}$, although the distribution is highly skewed. We therefore characterize the results using the empirical stability curve,
\begin{equation}
P(\Delta>x)=\frac{1}{N}\sum_{i=1}^{N}\mathbbm{1}(\Delta_i>x),
\end{equation}
where $\mathbbm{1}$ is the indicator function and $N$ is the total number of binaries. The right panel of Figure~\ref{fig:prob_fit} shows this distribution together with the best-fitting function,
\begin{equation}\label{eq:stype_fit}
\begin{split}
P(\Delta>x)=
C_1\Bigg[
&C_2\exp\!\left[-\left(\frac{x}{C_3}\right)^{C_4}\right] \\
&+(1-C_2)\exp\!\left[-\left(\frac{x}{C_5}\right)^{C_6}\right]
\Bigg],
\end{split}
\end{equation}
where $C_1=0.545$, $C_2=0.625$, $C_3=0.192$, $C_4=0.626$, $C_5=0.864$, and $C_6=4.052$. The fit reproduces the empirical distribution well except for the sharp decline near $\Delta\gtrsim0.95$, where 7.6\% of binaries experience extreme phase-space reductions caused by large encounter-induced eccentricity growth.

Evaluating the fit at $x=0.50$ gives $P(\Delta>0.50)\approx0.238$, implying that roughly 24\% of binaries originating from our modeled OC population experienced at least a factor of two reduction in the area available for stable coplanar circumstellar orbits while they were embedded in their natal OC.  Although Equation~\ref{eq:stype_fit} is derived from the full simulated binary population, it can be applied to any solar-type binary in the solar neighborhood to estimate the probability of a given reduction in stable circumstellar phase space from stellar encounters in an OC. 

If we assume the semi-major axis distribution of circumstellar planets is uniform, $\Delta$ is a good approximation of how much the primordial occurrence rates may be suppressed. However, the semi-major axis distribution likely depends on planet type. For example, rocky planets may preferentially form closer in to their host star and will be less affected than cold Neptunes and cold giant planets.  Therefore, we leave our results generalized in terms of stable circumstellar phase space so that they may be applied to any planet class as our understanding of formation and evolution for the various planet classes in binaries grow.

Combined with theoretical models of planet formation and the initial orbital distribution of circumstellar planets, Eq.~\ref{eq:stype_fit} provides a framework for estimating the fraction of primordial planets that are driven into dynamically unstable regions by OC evolution.  We further note that these results apply to initially coplanar configurations and do not include additional instabilities that may arise for inclined planetary orbits, such as destabilizing von Zeipel–Lidov–Kozai oscillations. Thus, our predictions represent an upper limit on the stable coplanar circumstellar phase space.

\section{Summary}\label{sec:conclusions}

We combine \textit{Gaia} DR3 observations, analytic and empirical models of OC formation and dissolution, and direct $N$-body simulations to quantify the contribution of dynamically processed OCs to the solar neighborhood binary population and assess the implications for primordial planet stability in these systems.  Using a synthetic population of OCs that formed over the past $5\,\rm Gyr$, we find that at least $53\%$ of FGK field binaries within $1\,\rm kpc$ originated in OCs with initial masses $\gtrsim200\,M_\odot$, with most arising from clusters that have since dissolved.

For CBPs, we estimate binary--binary encounter rates and give an empirical fit for the probability of a CBP destabilizing stellar encounter for a binary with separation $a_{\rm b}$ (Eq.\,\ref{eq:fit}).  For wider binaries ($a_{\rm b}\gtrsim100\,\rm au$), we quantify how stellar encounters reduce the stable circumstellar region using the \cite{Holman1999} stability criterion. We provide an empirical fit describing the distribution of co-planar circumstellar phase-space reduction (Eq.\,\ref{eq:stype_fit}), enabling predictions of how cluster dynamics suppress primordial circumstellar planet occurrence rates. Our predictions concern the stability of primordial planets with respect to their parent binaries and do not describe subsequent planet-level pathways such as scattering, ejection, exchange, or capture.

These results quantify the expected dynamical influence of OC evolution on binary planetary systems.  The encounter probabilities for individual binaries are generally modest, typically below 20\% over the binary separations considered. Nevertheless, because most nearby binaries likely originated in dynamically processed OCs, these encounter probabilities apply to a substantial fraction of the nearby binary population.  Consequently, planet occurrence rates in binaries should be interpreted within the context of their dynamical histories, providing a direct connection between cluster evolution and observed exoplanet demographics.

\begin{acknowledgments}
We thank the anonymous referee for their time and expertise in reviewing this work. Their comments and suggestions have helped improve the manuscript.  ACC and AMG acknowledge support from the National Science Foundation (NSF) under grant No. AST-2107738. Any
opinions, findings, and conclusions or recommendations expressed in this material are those of the author(s) and do not necessarily reflect the views of the NSF. This research was supported in part through the computational resources and staff contributions provided for the Quest high performance computing facility at Northwestern University which is jointly supported by the Office of the Provost, the Office for Research, and Northwestern University Information Technology. Publication of this work was supported by the AAS Publication Support Fund.
\end{acknowledgments}

%


\software{\texttt{nbody6++GPU} \citep{Wang2015code} \url{https://github.com/nbodyx/Nbody6ppGPU}, \texttt{scipy cKDTree} \citep{SciPy, ckdtree199}}




\bibliography{main}{}
\bibliographystyle{aasjournalv7.1}



\end{document}